\documentclass[preprint,aps,eqsecnum]{revtex4}
\usepackage{amsfonts}
\usepackage{graphics}
\usepackage{graphicx}
\usepackage{epsf}
\newcommand{\g}{\gamma}

\newcommand{\be}{\begin{equation}}
\newcommand{\ee}{\end{equation}}
\newcommand{\bea}{\begin{eqnarray}}
\newcommand{\eea}{\end{eqnarray}}

\begin{document}

\title{The Effective Potential in the Noncommutative Field Theories within the Coherent States Approach}
\author{M. A. Anacleto, J. R. Nascimento, A. Yu. Petrov}
\affiliation{Departamento de F\'\i sica, Universidade Federal da Para\'\i ba\\
Caixa Postal 5008, 58051-970, Jo\~ao Pessoa, Para\'\i ba, Brazil}
\email{anacleto, jroberto, petrov@fisica.ufpb.br}
\date{\today}

\begin{abstract}
We apply the coherent state approach to study the noncommutative scalar field theory with $\phi^4$ self-interaction and Yukawa coupling to the spinor field.
We verify that, contrarily to the commutative result, the scattering amplitude is ultraviolet finite. However,
the $\frac{1}{\theta}$ singularities arise as the noncommutative parameter $\theta$ tends to zero. For a special relation between two couplings, these singularities are shown to be cancelled, partially in the massive case and totally in the massless case.
\end{abstract}

\maketitle
\section{Introduction}
The noncommutativity of the space-time is considered now as its fundamental property \cite{SW}. In the present time there exist several manners of introducing the noncommutativity into the field theories. The most known one among them is the Moyal product formalism used in many papers. However, it is well known that the theories constructed on the base of the Moyal product display the ultraviolet/infrared (UV/IR) mixing generating the new class of infrared divergences which could break the perturbative expansion \cite{Minw}. This fact enforces interest to the alternative manners of introduction of the noncommutativity in the field theories which would allow to avoid such a difficulty (see discussion of different formulations of noncommutativity in \cite{SJ,Bal}). One of the efficient alternative approaches to the noncommutative field theories is the coherent state approach developed in \cite{Spallucci}. It was argued in \cite{Spallucci2} that applying of this formalism for the relativistic field theories leads to unitary and Lorentz invariant field theory whereas the Moyal product is known to break Lorentz invariance. Therefore the natural question consists in studying of general structure of quantum corrections in relativistic field theories within the coherent state approach. One more advantage of this approach consists in the fact that the UV/IR mixing is not produced within it, the only possible type of singularities are those ones arising for small values of the noncommutativity parameter $\theta$ \cite{An,Fer}. The natural problem is to study the possible types of singularities which could arise in the relativistic field theories formulated within this approach and to answer the question whether any extensions of the theories allowing complete or partial cancellation of the singularities could exist.  

We start with introduction of the noncommutative space-time. Its coordinates are defined to satisfy commutation relations
\begin{equation}
\label{noncom}
[\hat{x}^{\mu},\hat{x}^{\nu}]=i\theta^{\mu\nu}  \quad \quad \mu,\nu=0,...,D
\end{equation}
where $\theta^{\mu\nu}$, is an antisymmetric Lorentz tensor, which for even space-time dimension can be represented in terms of a
block diagonal form
\begin{equation}
\label{diag}
\theta^{\mu\nu}=diag(\hat{\theta}_{1},\hat{\theta}_{2},...,\hat{\theta}_{D/2}) \quad \quad \mu,\nu=0,...,D
\end{equation}
with 
\begin{equation}
\hat{\theta}_{i}=\theta_{i}\left(\begin{array}{clcr} 0 && 1 \\ -1 && 0 \end{array}\right).
\end{equation}
In the case of odd dimensional manifold the $(D,D)$-th diagonal element of the matrix is zero, and other diagonal blocks reproduce the structure (\ref{diag}). In other words, for the relativistic covariance of (\ref{noncom}) 
the spacetime can be foliated into noncommutative planes, defined by (\ref{diag}). Thus, conditions of Lorentz invariance and unitarity for an arbitrary field theory would imply that noncommutativity does not privilege any of such planes which is ensured by assuming a unique noncommutative parameter $\theta_{1}=\theta_{2}=...\theta_{D/2}=\theta$ in the theory. 

The coherent states approach is introduced to be an alternative one to the Moyal product approach. Indeed, within the Moyal product approach the propagators turn out to be unchanged whereas the vertices turn out to involve phase factors which presence is a main reason for the UV/IR mixing. In the coherent states approach the situation is opposite -- the noncommutativity (and the consequent nonlocality) is introduced in the way providing modification of the propagators by the Gaussian factor, therefore it is natural to suggest that vertices persist to be local in the space-time, which implies that they are unmodified under this approach \cite{Spallucci,Spallucci2,An,Fer}. We have showed in \cite{An} that these two approaches give compatible results in the case of the one-loop finite field theory, thus validity of the coherent states approach is confirmed. Also, it should be noted that though these approaches are compatible they are not equivalent since there is no local transformation of fields allowing to convert Moyal product formulation to the coherent state formulation. Indeed, the equality of the S-matrices is achieved only if the local transformations between two formulations of the field theory exist \cite{Lam}. 

Due to the Gaussian factors in the propagators all perturbative contributions turn out to be UV finite \cite{An}. In this paper, within the framework of this approach we find the one-loop contributions to the two- and four-point functions of the scalar field in the model involving $\lambda\phi^4$ and Yukawa couplings and study the effective potencial in this model. We find that unlike of the Moyal product formulation \cite{ourWZ}, within the framework of this formalism already the one-loop contribution to the effective potential is affected by the noncommutativity, with the leading, $\frac{1}{\theta}$ singularities turn out to be cancelled for a certain relation between couplings.

Similarly to \cite{An}, we can prove that the noncommutative corrections to the terms whose commutative counterparts are UV finite display exponential decay as distance grows. Indeed, the typical contributions arising within this approach after integration in the internal momentum $k$ but before of the expansion in power series in $\theta$, display the behaviour
\begin{eqnarray}
f(p)=\frac{e^{-\theta p^2}}{(p^2+m^2)^a},
\end{eqnarray}
where $p$ is the external momentum, and $a$ is a some integer number. Let us transform this expression to the coordinate space. We get
\begin{eqnarray}
f(x)=\int\frac{d^4p}{(2\pi)^4}\frac{e^{-\theta p^2+ipx}}{(p^2+m^2)^a},
\end{eqnarray}
which after integration, for $m\to 0$, in the case $a>0$, where the $\theta\to 0$ limit is not singular, gives
\begin{eqnarray}
f(x,\theta)\simeq f(x,\theta=0)-\frac{1}{16\pi^2}\frac{1}{(x^2)^{2-a}}e^{-\frac{x^2}{4\theta}}.
\end{eqnarray}
We see that in this case the $\theta$ dependent term representing itself as a noncommutative correction to the commutative counterpart $f(x,\theta=0)$, first, vanishes at $\theta=0$, second, exponentially decreases as distance grows. So we conclude that the noncommutative corrections are highly suppressed with the distance. If we have $a=0$ or $a=-1$, where the $\theta\to 0$ limit is singular
(the case $a\leq -2$ does not occur because of the general structure of the quantum corrections: indeed, $a=0$ case corresponds to logarithmic UV divergences of the commutative counterpart of the theory, whereas $a=-1$ -- to the quadratic ones), we have
\begin{eqnarray}
f_{a=0}(x)=\frac{1}{16\pi^2\theta^2}e^{-\frac{x^2}{4\theta}};\quad\, f_{a=-1}(x)=\frac{1}{16\pi^2\theta^2}(\frac{2}{\theta}-\frac{x^2}{4\theta^2})e^{-\frac{x^2}{4\theta}},
\end{eqnarray} 
so both these cases display exponential decay. As a result, we conclude that the noncommutative contributions are exponentially suppressed with the distance.

The paper is organized as follows. Section 2 is devoted to the calculation of the one-loop contribution to the two- and four-point vertex functions generated by the self-interaction of the scalar field. In the Section 3, the one-loop contribution to the two- and four-point vertex functions of the scalar field generated by the Yukawa coupling are found. The Section 4 contains calculation of the effective potential. In the Summary, the results are discussed.

\section{Two- and four-point functions of the scalar field for the $\lambda\phi^{4}$ coupling}

We start with applying the coherent state approach for $\lambda\phi^{4}$ theory. 
Its Lagrangian in Euclidean space is given by 
\begin{equation}
\label{f40}
{\cal{L}}_{sc}=\frac{1}{2}\partial_{\mu}\phi\partial^{\mu}\phi + \frac{m^{2}}{2}\phi^{2} + \frac{\lambda}{4!}\phi^{4}.
\end{equation}
Here the index $sc$ denotes that the model contains only scalar self-interaction.

According to the general formulation of the coherent state approach \cite{Spallucci,Spallucci2,An}, the propagator is augmented by the Gaussian factor:
\begin{equation}
D(p)=-\frac{e^{-\theta p^{2}/2}}{p^{2}+m^{2}},
\end{equation}
whereas suggestion of locality of vertex requires that it should remain to be unchanged (cf. \cite{Spallucci2,An,Fer})
\begin{equation}
\label{f4}
V(\phi)=\frac{\lambda}{4!}\phi^{4}.
\end{equation}

First we will consider the one loop correction to the mass term in the noncommutative $\phi^{4}$ theory. For this, we will calculate
the one-particle irreducible two-point function
\begin{equation}
\Pi(p)=<\tilde{\phi}(p)\tilde{\phi}(-p)>_{1PI}.
\end{equation}
In the one-loop approximation, this function is contributed by the "tadpole" diagram depicted at Fig. 1. Its contribution is given by 
\begin{equation}
\Gamma^{(2)}_{sc}=-\frac{\lambda}{2}\int\frac{d^{4}k}{(2\pi)^{4}}\frac{e^{-\theta k^{2}/2}}{k^{2}+m^{2}}.
\end{equation}
We introduce the standard Schwinger parametrization
\begin{equation}
\frac{1}{k^{2}+m^{2}}=\int^{\infty}_{0}d\alpha \, e^{-\alpha(k^{2}+m^{2})},
\end{equation}
and after doing the Gaussian momentum interation we obtain
\begin{equation}
\Gamma^{(2)}_{sc}=-\frac{\lambda}{32\pi^{2}}\int^{\infty}_{0}d\alpha \quad \frac{e^{-\alpha m^{2}}}{(\alpha + \frac{\theta}{2})^{2}}
=\frac{\lambda e^{\theta m^{2}/2}}{16\pi^{2}}\int^{\infty}_{\theta/2}ds \quad \frac{e^{-sm^{2}}}{s^{2}},
\end{equation}
where we have carried out a change of variables $s=\alpha + \theta/2$. Using the result 
\begin{equation}
\label{Exp}
E_{i}(\alpha x)=\int^{\infty}_{x}dt\frac{e^{-\alpha t}}{t}=-\gamma-\ln(\alpha x)-\sum^{\infty}_{n=1}\frac{(-1)^{n}(\alpha x)^{n}}{nn!},
\end{equation}
where the $\gamma$ is the Euler-Mascheroni constant, we get
\begin{equation}
\Gamma^{(2)}_{sc}=-\frac{\lambda}{32\pi^{2}}\left[ \frac{2}{\theta}-e^{\theta m^{2}/2}m^{2}
\left(\gamma+ \ln\left(\frac{\theta m^{2}}{2}\right) + \sum^{\infty}_{n=1}\frac{(-1)^{n}(\theta m^{2}/2)^{n}}{nn!}\right)\right].
\end{equation}
For small $\theta$, we have the following $\theta$ expansion of this contribution:
\begin{equation}
\label{2psc}
\Gamma^{(2)}_{sc}=-\frac{\lambda}{32\pi^{2}}\left[\frac{2}{\theta}-m^{2}\gamma(1 + \frac{\theta m^2}{2})- m^{2}\ln\left(\frac{\theta m^{2}}{2}\right) - \frac{\theta m^4}{2}\right] + O(\theta^2).
\end{equation}  
For $\theta=2/\Lambda^{2}$, with $\Lambda^{2}$ an ultraviolet cutoff $(\Lambda\rightarrow\infty)$, the well known result for the commutative analog of this theory is reproduced. 

Now let us study the four-point vertex function of the scalar field contributed by the four-leg supergraph depicted at Fig. 1. The corresponding analytical expression is 
\begin{equation}
\Gamma^{(4)}_{sc}=\frac{\lambda^{2}}{2}\int\frac{d^4k}{(2\pi)^{4}}\frac{e^{-\frac{\theta}{2}[k^2+(k-p)^2]}}
{(k^2+m^2)[(k-p)^2+m^2]},
\end{equation}
where $p=p_{1}+p_{2}$. Now we use the Schwinger representation 
\begin{equation}
\label{repres}
\frac{e^{-\frac{\theta k^2}{2}}}{k^2 + m^2}=e^{\theta m^2/2}\int_{\theta/2}^{\infty}ds e^{-s(k^2 + m^2)}.
\end{equation}
With prescription above the amplitude (\ref{repres}) becomes
\begin{equation}
\Gamma^{(4)}_{sc}=\frac{\lambda^2 e^{\theta m^2}}{2}\int_{\theta/2}^{\infty}ds\int_{\theta/2}^{\infty}dr\int\frac{d^4k}{(2\pi)^4}\exp[-s(k^2 + m^2)-r((k-p)^2 + m^2)].
\end{equation}
After performing the $k$ integration we have
\begin{equation}
\Gamma^{(4)}_{sc}=\frac{\lambda^2 e^{\theta m^2}}{32\pi^2}\int_{\theta/2}^{\infty}ds\int_{\theta/2}^{\infty}dr
\exp\left[-(s + r)m^2 + p^2\frac{sr}{s + r}\right].
\end{equation}
Now we introduce a change of variables $s=(1-x)\rho$ and $r=x\rho$. Thus,
\begin{equation}
\Gamma^{(4)}_{sc}=\frac{\lambda^2 e^{\theta m^2}}{32\pi^2}\int_{0}^{1}dx\int_{\theta(m^2-p^2x(1-x))}^{\infty}\frac{d\rho}{\rho}
e^{-\rho}.
\end{equation}
Using the result (\ref{Exp}), we obtain
\begin{eqnarray}
\Gamma^{(4)}_{sc}&=&-\frac{\lambda^2 e^{\theta m^2}}{32\pi^2}\left\{\gamma + \int_{0}^{1}dx\ln[\theta(m^2-p^2x(1-x))]\right. 
\nonumber\\
&&+ \left.\sum^{\infty}_{n=1}\frac{(-1)^{n}}{nn!}\int_{0}^{1}dx\theta^n(m^2-p^2x(1-x))^n  \right\}.
\end{eqnarray}
For small $\theta$, that is, $\theta m^2\ll 1$, $\theta p^2\ll 1$ this expression is expanded in $\theta$ as
\begin{eqnarray}
\Gamma^{(4)}_{sc}&=&-\frac{\lambda^2}{32\pi^2}\left\{\gamma(1+\theta m^2) + \ln(\theta m^2) -\theta \left(m^2 -\frac{p^2}{6}\right)\right. 
\nonumber\\
&&+ \left.(1 + \theta m^2) \int_{0}^{1}dx\ln\left[1-\frac{p^2x(1-x)}{m^2}\right]\right\} + O(\theta^2). 
\end{eqnarray}
Again the result for the commutative analog is reproduced for $\theta=2/\Lambda^{2}$. We find that the $\frac{1}{\theta}$ singularities arise in the theory. Their presence is caused by the fact that the exponential factor in the propagator plays the role of a natural regulator for the UV divergences, thus in the $\theta\to 0$ limit the singular behaviour of the quantum contributions is recovered. It is straightforward to show that the one-loop graphs with six and more legs do not display small $\theta$ singularities.

\section{Two- and four-point functions of the scalar field for the Yukawa coupling}
Let us consider the Yukawa model whose Lagrangian in the Euclidean space looks like
\begin{equation}
{\cal{L}}_{sp}=\frac{1}{2}\partial_{\mu}\phi\partial^{\mu}\phi + \frac{m^{2}}{2}\phi^{2} 
+ \bar{\psi}(\partial\!\!\!/ +iM)\psi +ih\bar{\psi}\psi\phi.
\end{equation}
Here the index $sp$ denotes presence of the coupling of scalar fielde to the spinor field.

The Feynman rules for the theory look as follows.
The propagator of spinor fields is
\begin{equation}
S(p)=\frac{i(p\!\!\!/-M)}{p^2+M^2}e^{-\theta p^2/2},
\end{equation} 
and the vertex is
\begin{equation}
\label{yu}
V(\phi;\psi)=ih\phi\bar{\psi}\psi.
\end{equation}
Two-point vertex function of the scalar field corresponding to the two-leg supergraph depicted at Fig. 2 turns out to have the following form:
\begin{eqnarray}
\Gamma^{(2)}_{sp}&=&-h^2\int\frac{d^4k}{(2\pi)^4}\frac{Tr[({k\!\!\!/}-M)(k\!\!\!/-p\!\!\!/-M)]}{(k^2+M^2)[(k-p)^2+M^2]}e^{-\frac{\theta k^2}{2}-\frac{\theta(k-p)^2}{2}}.
\end{eqnarray}
Using the modified Schwinger representation (\ref{repres}) we have
\begin{eqnarray}
\Gamma^{(2)}_{sp}&=&4h^2e^{\theta M^2}\int_{\theta/2}^{\infty}\int_{\theta/2}^{\infty} dsdr \int\frac{d^4k}{(2\pi)^4}[k\cdot(k-p)+M^2]
e^{-s(k^2+M^2)-r[(k-p)^2+M^2]}
\nonumber\\
&=&4h^2e^{\theta M^2}\int_{\theta/2}^{\infty}\int_{\theta/2}^{\infty} dsdr \int\frac{d^4k}{(2\pi)^4}\left[k^2+ \frac{r^2p^2}{(s+r)}+M^2\right]
\nonumber\\
&\times&{\mbox{exp}}{\left[-(s+r)k^2-\frac{rsp^2}{(s+r)}-(s+r)M^2\right]}.
\end{eqnarray}
After doing the internal momentum integration, we have 
\begin{eqnarray}
\!\!\!\!\!\!\!\!\!\Gamma^{(2)}_{sp}&=&\frac{h^2}{4\pi^2}e^{\theta M^2}\int_{\theta/2}^{\infty}\int_{\theta/2}^{\infty} dsdr \left[\frac{2+r^2p^2}{(s+r)^3} + \frac{M^2}{(s+r)^2}\right]\nonumber\\&\times&
{\mbox{exp}}\left[-\frac{rsp^2}{(s+r)} -(s+r)M^2\right].
\end{eqnarray}
Now we introduce a change of variables $s=(1-x)\rho$ and $r=x\rho$. Thus
\begin{eqnarray}
\Gamma^{(2)}_{sp}&=&\frac{h^2}{4\pi^2}e^{\theta M^2}\int_{0}^{1}dx\int_{\theta}^{\infty} d\rho \left[\frac{2}{\rho^2} + p^2x^2 + \frac{M^2}{\rho}\right]e^{-\rho f}
\nonumber\\
&=&-\frac{h^2}{4\pi^2}e^{\theta M^2}\int_{0}^{1}dx\left[M^2Ei[\theta f]+p^2x^2e^{-\theta f} 
+\frac{2e^{-\theta f}}{\theta}+2f\,Ei[\theta f]\right].
\end{eqnarray}
where $f=(x(1-x)p^2 + M^2)$. For small $\theta$, that is, $\theta m^2\ll 1$, $\theta p^2\ll 1$ this function is expanded as
\begin{eqnarray}
\label{2psp}
\Gamma^{(2)}_{sp}&=&\frac{h^2}{4\pi^2}\left[\frac{1}{\theta}-3M^2\gamma(1+ \theta M^2) -3M^2\ln(\theta M^2)
-\frac{\theta p^4}{20} \right.
\nonumber\\
&&-\left. \int_{0}^{1}dx[3M^2 + 2x(1-x)p^2](1 + \theta M^2)\ln\left(1+\frac{x(1-x)p^2}{M^2}\right)\right] + O(\theta^2).
\end{eqnarray}

We see that the contributions to the two-point function (\ref{2psc},\ref{2psp}) are UV finite. This situation differs not only from the commutative analogs of these  theories but also from their Moyal product based noncommutative formulations which display presence of the UV divergences (see f.e. \cite{Ar}), with in the last case, such divergences are generated by the planar parts of the contributions. Also we note that the leading, $\frac{1}{\theta}$ singularities in the two-point functions of the scalar field (\ref{2psp}) and (\ref{2psc}) turn out to be cancelled for $\lambda=4h^2$. This cancellation can be treated as a some analog of famous "miraculous cancellations" in supersymmetric field theories, despite actually the theory involving both $\lambda\phi^4$ (\ref{f4}) and Yukawa (\ref{yu}) couplings is not supersymmetric since numbers of bosonic and fermionic degrees of freedom in it are not equal.

\section{Effective potential}

Now our aim consists in calculation of the effective potential in the complete theory involving both $\lambda\phi^4$ and Yukawa couplings:
\begin{equation}
\label{acti}
{\cal{L}}_t=\frac{1}{2}\partial_{\mu}\phi\partial^{\mu}\phi + \frac{m^{2}}{2}\phi^2 
+ \bar{\psi}[\partial\!\!\!/ +i(M+h\phi)]\psi+\frac{\lambda}{4!}\phi^4.
\end{equation}
In this theory, one-loop effective potential receives contributions both from bosonic and fermionic sectors. It can be represented itself as a sum of two separate contributions: first of them is generated of all purely scalar loops, and other is a sum of all purely spinor loops with there is no mixed contributions.

To calculate the effective potential we follow the loop expansion method described in the book \cite{BOS}. First of all, we split the scalar field $\phi$ into sum of constant classical background field (mean field) $\Phi$ and quantum field $\chi$. The spinor field $\psi$ is suggested to be purely quantum one since we are interested only in the scalar sector of the effective action. It is known \cite{BOS} that the one-loop contribution to the effective action is totally determined by the quadratic action of quantum fields $\chi,\psi$ which looks like
\begin{equation}
\label{acti10}
{\cal{L}}_t[\Phi;\chi,\psi]=-\frac{1}{2}\chi\left(\Box-(m^2+\frac{1}{2}\lambda\Phi^2)\right)\chi  
+ \bar{\psi}\Big(\partial\!\!\!/+i(M+h\Phi)\Big)\psi .
\end{equation}

The one-loop effective Lagrangian $L^{(1)}(\Phi)$ (which in the constant field limit is reduced to the effective potential $U^{(1)}(\Phi)$) can be obtained from the generating functional \cite{BOS}: 
\bea
e^{iL^{(1)}(\Phi)}=\int D\chi D\psi\exp \left(i{\cal{L}}_t[\Phi;\chi,\psi]\right).
\eea
It is evident that the $U^{(1)}(\Phi)$ is a sum of two terms: the first of them which has been generated from the $\chi$ sector is equal to
\bea
\label{usc}
U_{sc}=\frac{i}{2}{\rm Tr}\ln\Big(\Box-(m^2+\frac{1}{2}\lambda\Phi^2)\Big),
\eea
and the second one which has been generated from the $\psi$ sector is equal to
\bea
\label{usp}
U_{sp}=-i{\rm Tr}\ln\Big(\partial\!\!\!/+i(M+h\Phi)\Big).
\eea

First we address to the scalar sector. The corresponding part of the effective potential can be graphically represented by the set of supergraphs depicted at Fig.1. Their sum is equal to
\bea
\label{effpots}
U_{sc}=\int\frac{d^4k}{(2\pi)^4}
\sum_{n=1}^{\infty}\frac{1}{2n}
\Big(-\frac{(2m^2+\lambda\Phi^2)e^{-\frac{\theta}{2} k^2}}{2k^2}\Big)^n,
\eea
The subscript $sc$ denotes origination of these contributions from the scalar part. This sum also can be obtained by expansion of the trace of the logarithm (\ref{usc}) in power series in $(m^2+\frac{1}{2}\lambda\Phi^2)$, with the $\Box^{-1}$ factors, following the prescriptions of the coherent states method, are replaced by 
\bea
-\frac{e^{-\frac{\theta}{2}k^2}}{k^2}.
\eea
After direct summation by the rule
\bea
\sum_{n=1}^{\infty}\frac{a^n}{n}=-\ln(1-a),
\eea
we get
\bea
U_{sc}&=&-\frac{1}{2}\int\frac{d^4k}{(2\pi)^4}\ln(1+\frac{(2m^2+\lambda\Phi^2)e^{-\frac{\theta}{2} k^2}}{2k^2}).
\eea
After replacement $k^2=2t$ we get
\bea
\label{intpots}
U_{sc}&=& -\frac{1}{8\pi^2}\int_0^{\infty}dt\, t
\ln[1+\frac{1}{4t}(2m^2+\lambda\Phi^2)e^{-\theta t} ].
 \eea
This integral is evidently finite. Unfortunately, it cannot be found in a closed form. However, its value can be estimated by analogy with \cite{ourWZ} as follows.
We approximate the integrand by its low-energy asymptotics, 
$ t \ln[1+\frac{1}{4t}(2m^2+\lambda\Phi^2) ]$, for $t<\Lambda$ 
(we suggest that on this interval the exponential can be approximated by 1), 
and by its high-energy asymptotics, $\frac{1}{4}(2m^2+\lambda \Phi^2)e^{-\frac{\theta}{2}
t}$, for $t>\Lambda$. We choose $\Lambda=\frac{1}{\theta}$ since the $1/\sqrt{\theta}$ is a natural energy scale. Thus,
\bea
U_{sc}&=&-\frac{1}{8\pi^2}\Big( \int_0^{\Lambda}dt
t\ln[1+\frac{1}{t}(\frac{1}{2}m^2+\frac{1}{4}\lambda \Phi^2)]+\int_{\Lambda}^{\infty}
dt (\frac{1}{2}m^2+\frac{1}{4}\lambda \Phi^2)e^{-\frac{\theta}{2} t}\Big) .
\eea
Integration gives the following result (up to the additive
constants):
\bea
\label{esc}
U_{sc}&=&-\frac{1}{8\pi^2}\Big[\frac{1}{2\theta^2}\ln\left(1+\theta(\frac{m^2}{2}+\frac{1}{4}\lambda\Phi^2)\right)+\frac{1}{\theta}(\frac{1}{2}m^2+\frac{1}{4}\lambda\Phi^2)(\frac{1}{2}+\frac{1}{e})\nonumber\\&+&\frac{1}{2}(\frac{1}{2}m^2+\frac{1}{4}\lambda\Phi^2)^2\ln\Big(
\frac{\theta(\frac{1}{2}m^2+\frac{1}{4}\lambda\Phi^2)}{1+\theta(\frac{1}{2}m^2+\frac{1}{4}\lambda\Phi^2)}
\Big)
\Big],
\eea
whose expansion gives
\bea
U_{sc}&=&\frac{1}{8\pi^2}\Big[
\frac{1}{\theta}(1+\frac{1}{e})(\frac{1}{2}m^2+\frac{1}{4}\lambda\Phi^2)+\frac{1}{2}(\frac{1}{2}m^2+\frac{1}{4}\lambda\Phi^2)^2
\Big(\ln\big(\theta(\frac{1}{2}m^2+\frac{1}{4}\lambda\Phi^2)\big)-\frac{1}{2}\Big)\nonumber\\
&-&\frac{1}{3}\theta(\frac{1}{2}m^2+\frac{1}{4}\lambda\Phi^2)^3
\Big]+O(\theta^2)
\eea
We see that this part of the effective potential is linearly singular at $\theta=0$.
This is very natural since its commutative analog is UV divergent, 
and the noncommutativity plays the role of the UV regulator.

Now we calculate the spinor sector contribution to the effective potential. 
The one-loop effective potential is contributed from the spinor sector by cyclic diagrams formed by spinor propagators with
external lines are background scalar fields $M+h\Phi$. Such diagrams are depicted in Fig. 2. 
The total contribution from the spinor sector to the one-loop effective potential is
given by the expression 
\bea 
\label{effpot}
U_{sp}=-{\rm Tr}\int\frac{d^4k}{(2\pi)^4}\sum_{n=1}^{\infty}\frac{1}{n}\Big((M+h\Phi)\frac{e^{-\frac{\theta k^2}{2}}}{k^2}
(\g\cdot k)\Big)^n, 
\eea 
where $\frac{1}{n}$ is a symmetry factor, and $\Lambda(k)$ is given by (\ref{lambda}). The subscript $sp$ denotes that this effective potential emerges from the spinor sector. We also took into account that each fermionic loop carries minus sign. 
This sum also can be obtained by expansion of the trace of the logarithm (\ref{usp}) in power series in $M+h\Phi$, with the $\Box^{-1}$ factors, following the prescriptions of the coherent states method, are replaced by 
\bea
\label{lambda}
F(k)=\frac{e^{-\frac{\theta k^2}{2}}}{k^2}.
\eea
Then, it is easy to find that 
\bea 
(\g\cdot k)^{2l}=(-k^2)^l;\quad\, (\g\cdot
k)^{2l+1}=(-k^2)^l(\g\cdot k). 
\eea 
Since $F(k)$ is an even function of the momentum $k$, only even terms in (\ref{effpot})
will give nontrivial contributions. Therefore the $U_{sp}$ takes the form 
\bea
\label{effpot1}
U_{sp}=-4\int\frac{d^4k}{(2\pi)^4}\sum_{n=1}^{\infty}\frac{1}{2n}\Big(-(M+h\Phi)^2
F^2(k)k^2\Big)^n. 
\eea 
The factor 4 is caused by the trace of the unit $4\times 4$ matrix.
It remains to substitute the expression
for $F(k)$ (\ref{lambda}), so in the Euclidean space we get 
\bea
\label{effpot2}
U_{sp}=-4\int\frac{d^4k}{(2\pi)^4}\sum_{n=1}^{\infty}\frac{1}{2n}\Big(-(M+h\Phi)^2
\frac{e^{-\theta k^2}}{k^2}\Big)^n, 
\eea 
which after summation gives
\bea
U_{sp}&=&2\int\frac{d^4k}{(2\pi)^4}\ln[1+ (M+h\Phi)^2
\frac{e^{-\theta k^2}}{k^2}].
\eea
After replacement $k^2=t$ we get
\bea
\label{effpotspin}
U_{sp}&=& \frac{1}{8\pi^2}\int_0^{\infty}dt\, t
\ln[1+\frac{1}{t}(M+h\Phi)^2e^{-\theta t}].
\eea
This integral is evidently finite. Let us estimate its value as above. We approximate the integrand by its low-energy
asymptotics, $t
\ln[1+\frac{1}{t}(M+h\Phi)^2]$ for
$t<\Lambda$ (we suggest that on this interval we can approximate exponent by 1),
and by its high-energy asymptotics, $(m+h\Phi)^2 e^{-\theta t}$,
for $t>\Lambda$, with $\Lambda=\frac{1}{\theta}$. Thus, 
\bea
U_{sp}&=&\frac{1}{8\pi^2}\Big(\int_0^{\Lambda}dt
t\ln[1+\frac{1}{t}(M+h\Phi)^2]+\int_{\Lambda}^{\infty}
dt (M+h\Phi)^2 e^{-\theta t}\Big). 
\eea 
Integration gives the
following result (up to the additive constants): 
\bea
\label{esp}
U_{sp}&=&\frac{1}{8\pi^2}\Big[\frac{1}{2\theta^2}\ln(1+\theta(M+h\Phi)^2)+\frac{1}{2\theta}(M+h\Phi)^2\nonumber\\&+&\frac{1}{2}(M+h\Phi)^4\ln\Big(
\frac{\theta(M+h\Phi)^2}{1+\theta(M+h\Phi)^2}
\Big)+
\frac{(M+h\Phi)^2}{\theta}\frac{1}{e}\Big]. 
\eea 
After expansion in $\theta$ we get
\bea
U_{sp}&=&\frac{1}{8\pi^2}\Big[
\frac{(M+h\Phi)^2}{\theta}(1+\frac{1}{e})+\frac{1}{2}(M+h\Phi)^4\Big(\ln(\theta(M+h\Phi)^2)-\frac{1}{2}\Big)\nonumber\\
&-&\frac{1}{3}\theta(M+h\Phi)^6
\Big]+O(\theta^2).
\eea
We see that this part of
effective potential also displays linear singularity at $\theta=0$. This is very
natural since its commutative analog is UV divergent, and the weak
noncommutativity singularity is a consequence of regularization of the UV
divergence.

Now let us consider the total one-loop effective potential in the theory. This effective potential is evidently a sum of contributions from spinor and scalar sectors given by (\ref{esp}) and (\ref{esc}) respectively, and after omitting of constant terms it turns out to be equal to
\bea
\label{etot}
U_{whole}&=&\frac{1}{8\pi^2}\Big[\frac{1}{\theta}(2Mh\Phi+h^2\Phi^2)(1+\frac{1}{e})+\frac{1}{2}(M+h\Phi)^4(\ln(\theta(M+h\Phi)^2)-\frac{1}{2})\nonumber\\&-&
\frac{1}{3}\theta(M+h\Phi)^6-\nonumber\\&-&\frac{1}{4\theta}\lambda\Phi^2(1+\frac{1}{e})-\frac{1}{2}(\frac{1}{2}m^2+\frac{1}{4}\lambda\Phi^2)^2
\Big(\ln(\theta(\frac{1}{2}m^2+\frac{1}{4}\lambda\Phi^2))-\frac{1}{2}\Big)\nonumber\\
&+&\frac{1}{3}\theta(\frac{1}{2}m^2+\frac{1}{4}\lambda\Phi^2)^3
\Big]+
O(\theta^2).
\eea 
We find that the term proportional to $\frac{\Phi^2}{\theta}$ would disappear in the case of the special relation between couplings $h$ and $\lambda$, that is, 
$\lambda=4h^2$. At the same time, we remind that cancellation of $\frac{1}{\theta}$ singularities
in the contributions to two-point function (\ref{2psc},\ref{2psp}) yields the same relation between couplings (exactly the same condition for cancellation of the quadratic divergences arises in the commutative case). Moreover, in the case $M=m=0$ the complete one-loop effective potential $U_{whole}$ {\it turns out to be identically equal to zero independently of the calculation scheme}, i.e. the bosonic and fermionic sectors mutually cancel each other.
This cancellation supports the hope that the correct supersymmetric extension of the scalar field theory would contain no $\frac{1}{\theta}$ singularities, by analogy with the well known fact that the Wess-Zumino 
model contains only logarithmic divergences.

\section{Summary}
Let us summarize the results. We have shown that within the coherent states formalism the relativistic field theories are UV finite. Also, we demonstrated that the essential advantage of this method consists in preservation of Lorentz covariance whereas in the Moyal product approach there exists a preferred frame because of presence of the constant antisymmetric noncommutativity matrix which breaks Lorentz invariance. We demonstrated that within the coherent state approach the only type of singularities can arise, that is, the singularities of the weak noncommutativity limit. The important property of this formulation is that unlike of the Moyal product approach, within the coherent state formalism the quantum contributions do not include planar part which is known to generate UV divergences. 

We have calculated the effective potential in the theory involving scalar and spinor fields and found that there is a fundamental difference with the Moyal product based noncommutative field theories. Indeed, within the coherent state approach already first loop contribution to the effective potential is affected by the noncommutativity whereas within Moyal product approach the one-loop effective potential in the field theory coincides with its commutative analog \cite{ourWZ}.

The model under consideration is not supersymmetric since the numbers of bosonic and fermionic degrees of freedom are not equal. Indeed, after introducing the auxiliary field $F=\phi^2$ which can be eliminated via equations of motion similarly to the Wess-Zumino model, this theory contains two real bosonic degrees of freedom and two complex fermionic degress of freedom. However, despite of this,
at the certain relation of coupling constants the cancellation of the leading ($\frac{1}{\theta}$) divergences, partial for massive case and complete for massless case is achieved. This fact is based on the well-known phenomenon of the mutual cancellation of bosonic and fermionic contributions which can be observed in principle not only in the supersymmetric field models but in any models involving bosonic and fermionic fields, the similar situation takes place in the (noncommutative) $CP^{N-1}$ model \cite{cpn} where the partial cancellation of dangerous (linear) UV and UV/IR infrared divergences takes place in the model minimally coupled to fermions, whereas in the supersymmetric case they are completely removed. Therefore it is natural to expect that supersymmetrization of the theory would allow to eliminate totally the weak noncommutativity singularities  which presence is the natural consequence of the UV divergences of the corresponding commutative field theories.
In the paper we gave some support to this suggestion by demonstrating the partial cancellation of $\frac{1}{\theta}$ divergences in the model of scalar field minimally coupled to fermions for the certain relation between couplings in the massive case and exact vanishing of the one-loop effective potential for the same relation between couplings in the massless case. The next step consists in more detailed study of the problem of compatibility of the coherent state formalism with the supersymmetry. This will be done in the forthcoming paper.

\vspace*{3mm}

{\bf Acknowledgements.}
The authors would like to thank D. Bazeia, M. Gomes and F. Brandt for useful discussions. This work was partially supported by PRONEX/CNPq/FAPESQ. 
M. A. A. has been supported by CNPq/FAPESQ DCR program, project No. 350136/2005-0. A. Yu. P. has been 
supported by CNPq/FAPESQ DCR program, project No. 350400/2005-9.

\begin{figure}[ht]
\includegraphics{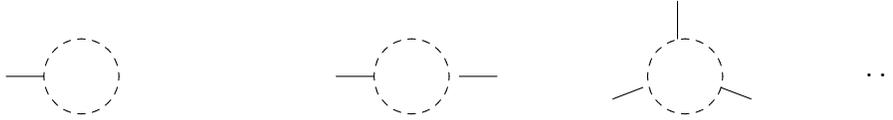}
\caption{One-loop effective potential in the Yukawa model.}
\label{Fig1}
\end{figure}

\begin{figure}[ht]
\includegraphics{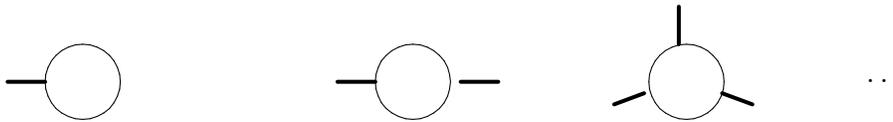}
\caption{One-loop effective potential in the $\phi^4$ model.}
\label{Fig2}
\end{figure}

\end{document}